**Full title:** Focal Loss Analysis of Nerve Fiber Layer Reflectance for Glaucoma Diagnosis

**Word count:** 5053

**Tables:** 6; **Figures:** 9

**Author:** Ou Tan, PhD[1] Liang Liu, MD,[1] Qisheng You, MD,[1] Jie Wang, MS[1] Aiyin Chen, MD[1] Eliesa Ing, MD[1] John C. Morrison, MD[1] Yali Jia, PhD[1] David Huang, MD, PhD[1]

[1]Casey Eye Institute, Oregon Health & Science University

***Corresponding Author**: David Huang, MD, PhD

Peterson Professor of Ophthalmology & Professor of Biomedical Engineering

Casey Eye Institute, Oregon Health & Science University

515 SW Campus Dr., CEI 3154

Portland, OR, USA 97239-4197

503-494-5131

Email: huangd@ohsu.edu

**Funding Sources:** NIH grants R01 EY023285, R21EY027007 and P30 EY010572, and an unrestricted grant from Research to Prevent Blindness to Casey Eye Institute.

**Commercial relationships:** OT: P; LL: None; QY: None; JW: None; AC: None; EI: None; JM: None; YJ: F, P, DH: F, R, I, P,

# Abstract

**Purpose**: To evaluate nerve fiber layer (NFL) reflectance for glaucoma diagnosis.

**Methods**: Participants were imaged with 4.5×4.5-mm volumetric disc scans using spectral-domain optical coherence tomography (OCT). The normalized NFL reflectance map was processed by an azimuthal filter to reduce directional reflectance bias due to variation of beam incidence angle. The peripapillary area of the map was divided into 160 superpixels. Average reflectance was the mean of superpixel reflectance. Low-reflectance superpixels were identified as those with NFL reflectance below the 5 percentile normative cutoff. Focal reflectance loss was measure by summing loss in low-reflectance superpixels.

**Results**: Thirty-five normal, 30 pre-perimetric and 35 perimetric glaucoma participants were enrolled. Azimuthal filtering improved the repeatability of the normalized NFL reflectance, as measured by the pooled superpixel standard deviation (SD), from 0.73 to 0.57 dB (p<0.001, paired t-test) and reduced the population SD from 2.14 to 1.78 dB (p<0.001, t-test). Most glaucomatous reflectance maps showed characteristic patterns of contiguous wedge or diffuse defects. Focal NFL reflectance loss had significantly higher diagnostic sensitivity than the best NFL thickness parameter (overall, inferior, or focal loss volume): 53% v. 23% (p=0.027) in PPG eyes and 100% v. 80% (p=0.023) in PG eyes, with the specificity fixed at 99%.

**Conclusions**: Azimuthal filtering reduces the variability of NFL reflectance measurements. Focal NFL reflectance loss has excellent glaucoma diagnostic accuracy compared to the standard NFL thickness parameters. The reflectance map may be useful for localizing NFL defects.

# 1 Introduction

Nerve fiber layer (NFL) thickness measurement by optical coherence tomography (OCT) has been widely used in the clinical management of glaucoma.[1-7] According to the American Academy of Ophthalmology's IRIS Registry, OCT is now used in more glaucoma clinic visits than visual field (VF) analysis.[8] NFL thickness is useful for confirming the diagnosis of glaucoma and monitoring the progression, at least in the early stage.[9-14] However, its diagnostic sensitivity is not sufficient to be used alone for mass screening.[10, 15] At the 99% specificity diagnostic cut-off needed for screening applications, the best single NFL thickness parameters have sensitivity of only 7-30% for pre-perimetric glaucoma (PPG) and 20-60% for perimetric glaucoma (PG).[16-21] Combining diagnostic parameters from several anatomic regions could boost the sensitivity to 55-85% for perimetric glaucoma. [16, 18, 22-26] Thus there is still room for improvement.

In this article, we explore methods of improving glaucoma diagnostic accuracy by analyzing the NFL reflectance. It is well established that the NFL reflectivity is reduced in glaucoma subjects,[27] presumably due to loss of axons and axonal microtubule content.[28-30] However, the average NFL reflectivity, as a diagnostic parameter, underperformed the average NFL thickness.[27] The diagnostic accuracy could be improved by normalizing the NFL reflectivity by one or more outer retinal layers, but the resulting average reflectivity ratio still does not outperform NFL thickness.[31, 32] Combining the reflectivity ratio with NFL thickness does improve diagnostic accuracy.[32, 33] Thus we use the cumulative normalized NFL reflectance (summing the reflectivity ratio over the thickness of the NFL) as the starting point of this analysis. Our aim is to further improve diagnostic accuracy by reducing measurement noise and accentuating focal loss.

We hypothesize that an important limitation of the diagnostic reliability of NFL reflectivity is the dependence on incidence angle. The NFL reflectivity is highly directional in the plane parallel to the long axis of the nerve fibers.[34, 35] In routine clinical OCT imaging, it is very difficult for the operator to obtain uniform perpendicular beam incidence that would maximize reflectivity. Generally, beam incidence angle varies within any NFL scan circle or area, leading to measurement variability that reduces diagnostic accuracy. In this study, we reduced this variability by performing azimuthal filtering of NFL reflectance in post-processing of the OCT images. In a further step, we developed an algorithm to quantify focal loss of NFL reflectance. We hypothesize that, compared to global averaging, focal loss is less sensitive to incidence angle and other measurement noise, and that it is more sensitive to the focal and regional patterns of damage common in early glaucoma. One goal of this approach was to improve diagnostic accuracy by combining these two approaches. Thus, the diagnostic performance of our novel NFL reflectance parameters is tested in this prospective observational clinical study.

# 2 Methods

## 2.1 Participants

This prospective observational study was performed from January 06, 2017 to May 30, 2019 at the Casey Eye Institute, Oregon Health & Science University (OHSU), Portland, OR, USA. The research protocols were approved by the Institutional Review Board at OHSU and carried out in accordance with the tenets of the Declaration of Helsinki. Written informed consent was obtained from each participant. The study was in accordance with the Health Insurance Portability and Accountability Act of 1996 (HIPAA) privacy and security regulations.

All participants were part of the "Functional and Structural Optical Coherence Tomography for Glaucoma" study (FSOCT, R01 EY023285). The inclusion criteria for the PG group were (1) an optic disc rim defect (thinning or notching) or retinal NFL defect visible on slit-lamp biomicroscopy, and (2) a consistent glaucomatous pattern on both qualifying Humphrey SITA 24-2 VFs. The consistent pattern was defined as either pattern standard deviation (PSD) outside normal limits ($p < 0.05$) or glaucoma hemifield test outside normal limits. Eyes in the PPG group met the biomicroscopic criteria, but not the VF criteria for the PG group.

For the normal group, the inclusion criteria were as follows: (1) No evidence of retinal pathology or glaucoma, (2) a normal Humphrey 24-2 VF, (3) intraocular pressure < 21 mm Hg, (4) central corneal pachymetry > 500 µm, (5) no chronic ocular or systemic corticosteroid use, (6) an open angle on gonioscopy, (7) a normal appearing optic nerve head (ONH) and NFL, and (8) a symmetric ONH between left and right eyes.

Participants were excluded from this study if any of the following situations were observed: (1) best-corrected visual acuity less than 20/40, (2) age < 40 or >80 years, (3) spherical equivalent refractive error of > +3.00D or < -7.00 diopters, (4) previous intraocular surgery except for an uncomplicated cataract extraction with posterior chamber intraocular lens implantation, (5) any other diseases that might cause VF loss or optic disc abnormalities, or (6) inability to perform reliably on automated VF testing.

One eye from each participant was scanned and analyzed. For the normal group, the eye was randomly selected. For the PPG and PG group, the eye with the worse VF mean deviation was selected.

## 2.2 Data Acquisition

Participants were scanned with a 70 kHz, 840 nm wavelength spectral-domain OCT system (Avanti, Optovue, Inc., Fremont, CA, USA). Two scan patterns, the optic disc volumetric high-definition OCT angiography (HD OCTA) scan and the structural OCT ONH scan, were used.

The optic disc volumetric HD OCTA scan covered 4.5×4.5 mm area centered on the disc. The cross-sectional B-frames, comprised of 400 A-lines, were repeated twice at each location to allow the computation of the angiographic flow signal.[36] Each volume was comprised of 400 B-frame locations. Two consecutive volumetric scans, i.e., a vertical-priority raster and a horizontal-priority raster, were merged using an orthogonal registration algorithm. This reduced motion artifacts and improved image quality.[36-38] The merged volume provided both angiographic (flow signal) and structural (reflectance signal) images. Volumetric structural OCT images were analyzed by our novel reflectance algorithm described below. Good quality images with a signal strength index (SSI) of 50 (out of 100) or more and a quality index of 5 (out of 10) or more were used. Images not meeting the quality criteria were excluded from further analysis.

The ONH scan was a 4.9 mm composite scan, centered on the disc. Using the Avanti software, the ONH scan provided the traditional NFL thickness profile and measurements on the circle with a diameter of 3.4 mm. Although we could obtain a similar thickness profile from the volumetric scan, we chose to use the traditional ONH scan because the diagnostic performance and quality control has been well characterized in the literature.[39, 40]

The VF was assessed by standard automated perimetry on the Humphrey Field Analyzer (HFA II; Carl Zeiss Meditec, Inc., Dublin, CA, USA), using the Swedish Interactive Thresholding Algorithm 24-2.

## 2.3 NFL Reflectance Analysis

### 2.3.1 Image segmentation

The OCT signal of the merged volumetric scan was exported from the Avanti and processed by the custom software Center for Ophthalmic Optics & Lasers-Angiography Reading Toolkit (COOL-ART) that was developed in our laboratory in the MATLAB programming environment by coauthors YJ, JW, and others.[41] COOL-ART automatically segmented the disc boundary and retinal layers and allowed manual correction by human graders. Grading was conducted by co-authors LL and QY.

### 2.3.2 Normalized NFL Reflectance Map

The NFL reflectance (Fig. 1) was analyzed using custom software developed by the first author (OT). The OCT reflectance data were transformed to a linear intensity scale. The NFL band and the photoreceptor and pigment epithelium complex (PPEC) band were extracted from the OCT image. The PPEC band included the region from the anterior boundary of ellipsoid zone (EZ) to the Bruch's membrane. The OCT intensity was axially averaged in the PPEC band to provide a reference map. The NFL reflectance was axially summed to provide the NFL reflectance map (Fig. 1 A-C). Based on the data from normal subjects, the NFL/PPEC reflectance ratio map was normalized by the population average of map averages in the 1.1 – 2.0 mm radius analytic zone, followed by transformation to a logarithmic dB scale. For the sake of brevity, we refer to this output as the NFL reflectance map. Because large vessels displace nerve fibers and interfere with NFL reflectance analysis,[42] the reflectance values in vessel areas were replaced with values from neighboring pixels to preserve continuity (Fig.1 D-F).

### 2.3.3 Polar Coordinate Spatial Frequency Filtering of the Reflectance Map

The NFL reflectance signal in an OCT image not only depends on the intrinsic reflectivity but also on extrinsic factors such as beam incidence angle and beam coupling factors. Generally, these extrinsic factors vary with the azimuthal angle, which is the angular position of the peripapillary retina in the polar coordinates. The origins of this variation are discussed below. To reduce the effects of the extrinsic factors, we performed azimuthal spatial frequency filtering. We first transformed the NFL reflectance map from Cartesian to polar coordinates, with the disc center as the origin. Then a notch filter was applied to the transverse spatial spectrum to remove the first-order sinusoidal component along the azimuthal dimension. The filter also removed high frequency components along both radial and azimuthal dimensions, a smoothing action that reduces speckle noise. The result accentuates nerve fiber bundle defects (Fig. 1G). The disc area was masked out because the NFL reflectance was undefined in this region.

### 2.3.4 Superpixel

The filtered NFL reflectance map was divided into superpixels (Fig. 1H). The superpixel grid in the peripapillary area was divided into 32 tracks that ran parallel to the average nerve fiber trajectory map determined by nerve fiber flux analysis described in a previous publication.[43] The widths of the tracks were adjusted so that each contained the same nerve fiber flux, defined by the cross-sectional area of the NFL that was cut perpendicular to nerve fiber trajectory. Thus each track contained approximately an equal number of nerve fibers. Because the NFL is thicker at the superior and inferior arcuate bundle regions, the tracks there were narrower. Thus the arcuate regions were weighed more densely with superpixels, which is appropriate as these regions are more likely to be affected by glaucoma. Each track was evenly divided into 5 segments in the annulus between 1.1 and 2.0 mm from the center of the disc. The region outside of the 2.0-mm radius was excluded to avoid cropping artifacts from possible scan decentration. Thirty-two tracks in 5 segments resulted in 160 superpixels. The NFL reflectance in each superpixel was averaged. Experimentation with different sizes of superpixels resulted in little variation in diagnostic performance. Though the diagnostic performance would be slightly worse if the superpixel size was much larger or smaller.

### 2.3.5 Age, Gender and Axial Length Adjustment Using Linear Mixed Effects Model

Multiple linear regression based on the linear mixed effects model[44, 45] was used to test the correlation between age, gender and axial length and the normalized NFL reflectance in the normal group. The superpixel location was modeled as a random effect, while age, axial length, and gender, were used as fixed effects. Age, axial length, and the interaction between them were significant factors. Therefore, the NFL reflectance of superpixels were adjusted for age and axial length using the regression model obtained from normal eyes.

### 2.3.6 Low-Reflectance Superpixel

We assumed that the normalized NFL reflectance followed a normal distribution in the normal group. This was confirmed by the Shapiro-Wilk test (p=0.42). The population average and standard deviation of the adjusted NFL reflectance for each superpixel were calculated. Based on the normal distribution assumption, the 5% and 1% cutoff of reflectance values were estimated for each superpixel. Superpixels with adjusted reflectance below the 5% cutoff were considered "low-reflectance." The number of low-reflectance superpixels was counted for each eye.

### 2.3.7 Diagnostic Parameters

Besides the low-reflectance superpixel count, two additional diagnostic parameters were calculated: overall average reflectance and focal reflectance loss. The overall average reflectance was the average of reflectance values in all superpixels in an eye. Focal reflectance loss was the summation of reflectance deviation (difference between the tested superpixel and the normal reference adjusted for age and axial length) over the low-reflectance superpixels, then divided by the total number of superpixels (n = 160). Glaucoma damage manifests as more low-reflectance pixels (positive integer count), lower overall average reflectance (dB), and more negative focal reflectance loss (dB).

The above NFL reflectance parameters were compared with the two standard glaucoma diagnostic parameters already in clinical use: overall average of NFL thickness and VF mean deviation (MD). The overall NFL thickness and quadrant NFL thickness at the 3.4-mm diameter circumpapillary circle was obtained from the ONH scan using the REVue software (version 2018.0.0.18) provided by the manufacturer. The focal loss volume of NFL thickness was calculated based on the NFL thickness profile.[46]

### 2.3.8 Statistical Analysis

The two-sided Wilcoxon rank sum test was used to compare the difference between the normal and glaucoma groups. The diagnostic accuracy was evaluated by the area under receiving characteristic operating curve (AROC)[17] and by the sensitivity at the 99% specificity. The sensitivity was compared using McNemar's test. For all parameters, the age adjustment was applied to obtain equivalent value at a reference age of 50 years.[47] Pearson correlation coefficients were calculated among NFL parameters and VF MDs. The coefficients were compared using the bootstrap method.[48] All analysis were done in Matlab R2019a with statistics toolbox.

We used cross validation to reduce bias in the diagnostic accuracy measurement. We chose the 0.632+ bootstrap with replacement for the age and axial length adjustment, and low-reflectance cutoff calculations.[49-51] The parameters were averaged from multiple trials. In each trial the parameters were estimated based on 63.2% of normal population and applied to other normal and glaucoma eyes.

## 3 Results

### 3.1 Characteristics of the Study Participants

One eye each from normal (n=35), PPG (n=30), and PG participants (n = 35) were included in this study. Patients in both the PPG and PG groups were significantly older and had longer axial lengths, worse VF MDs, and worse PSDs than normal patients (Table 1). In the PPG group, MD ranged from -7.3 to 2.0 dB, and PSD from 1.1 to 4.0 dB. In the PG group, VF MD ranged from -19.3 to 0.3 dB, and PSD from 1.4 to 14.7 dB.

### 3.2 Incidence Angle and Azimuthal Filtering

Our azimuthal filtering method was based on the assumption that incidence angle variation along the circumpapillary circles has first-order sinusoid variation due to the nasal location of the disc relative to the optical axis of the eye. The phase and magnitude of this variation depends on the geometry of the eye as well as the position of the OCT scan beam relative to the pupil. Generally, the incidence angle is positive (centripetal) because the scan-mirror conjugate plane in the pupillary plane is anterior to the center of curvature of the retina (i.e., the retina appears concavely curved on an OCT cross-section). The retina nasal to the disc has the largest incidence angle if the OCT beam is centered in the pupil. In the normal group, the incident angle along the 3.4-mm diameter circle indeed showed the expected first-order azimuthal sinusoidal variation with amplitudes ranging from 1.5° to 9.6° with a constant offset of 2.8 to 11.1° (Fig. 2B). Due to the centripetal offset, the absolute value of the incidence angle also generally varied as a first-order sinusoid along the analytic circle around the disc, with an amplitude of 6.1±3.7°, significantly above zero (p<0.001). The normalized reflectance decreased when the absolute incidence angle increased (Fig. 2C). The amplitude of first-order azimuthal sinusoidal reflectance variation was significantly correlated with the first-order sinusoidal variation in absolute incidence angle variation (r=0.421, p=0.007). In contrast, the amplitudes of other azimuthal sinusoidal orders, i.e., 0, 2, 3, 4, were not significantly correlated (p>0.17). These results supported the premise behind azimuthal filtering to remove the first-order periodicity. We compared the diagnostic accuracy of NFL reflectance parameters obtained with and without azimuthal filtering and found that the accuracy of all parameters was always better with filtering on. Thus all of the following results were produced with the filter on.

There were also second-order sinusoidal variations in incidence angle (Fig. 2B) associated with normally thicker NFL locations superiorly and inferiorly. Our azimuthal filtering preserved this spatial frequency component as it contained diagnostic information.

Using 20 normal eyes with two repeated OCT scans, we tested the effect of azimuthal filtering on the repeatability of NFL reflectance in the 160 superpixels. The repeatability was measured by the pooled standard deviation (SD). For the superpixels, the repeatability was improved from 0.73±0.15 dB to 0.57±0.11 dB (p<0.001, paired t-test) using the azimuthal filter.

In the normal group with 35 eyes, we also compared the population SD for each superpixel. It was reduced from 2.14±0.40 dB to 1.78±0.34 dB using the azimuthal filter. The reduction was significant (p<0.001, paired t-test).

These improvement in repeatability and reduction in population variation showed that azimuthal filtering reduced the effects of incidence angle variation and other biases in NFL reflectance measurement.

### 3.3 Reflectance Patterns in Normal and Glaucoma Groups

The normalized NFL reflectance map, averaged in the normal group (Fig. 3), had the highest reflectance in the inferotemporal (6:30 o'clock peak, using right eye convention) and superotemporal (11 o'clock peak) regions. There was also a secondary superonasal (1 o'clock) peak. The population SD map showed slightly higher variability in the inferotemporal and superonasal regions. The average SD was 1.8 dB and the peak SD was 2.4 dB.

The average pattern of reflectance loss in the glaucoma groups (Fig. 3) showed that damage was commonly most severe in the inferotemporal region (7 o'clock peak), followed by shallower peaks superotemporally (11 o'clock) and superonasally (1:30 o'clock). The average loss was 2.2 dB in the PPG group and 5.6 dB in the PG group. The peak loss (inferotemporal) was 3.1 dB in the PPG group and 8.1 dB in the PG group.

Three eyes were selected from the normal, PPG, and PG groups to show the characteristic glaucomatous reflectance loss patterns (Fig. 4). Both PPG and PG eyes had wedge-shaped loss patterns consistent with the nerve fiber wedge defect characteristic of glaucoma. The reflectance loss pattern correlated well with the locations of VF defects.

We tested whether or not most NFL reflectance loss patterns were consistent with nerve fiber wedge defects characteristic of glaucoma. To perform this analysis, we categorized the loss pattern into diffuse, wedge, other grouping, isolated, and none (Fig. 5). Diffuse loss (full width defect spanning more than a quadrant of the annular analytic area) would be consistent with severe glaucoma, while wedge pattern (contiguous superpixels connecting the inner and outer edges of the annular analytic zone) would be consistent with mild or moderate glaucoma when damage was local. Reflectance loss in isolated superpixels or other grouping (3 or more contiguous superpixels in a non-wedge configuration) could indicate measurement noise or mild disease of indeterminate type. If two or more patterns were observed in same eye, the one corresponding to a more severe glaucoma category was applied.

There was a positive correlation between the eyes with severe defects and glaucoma stages. Most PPG eyes (22 of 30) exhibited glaucomatous reflectance loss patterns (Table 2), and all of the PG eyes exhibited glaucomatous (diffuse or wedge) reflectance loss patterns. Nineteen normal eyes exhibited isolated or other-grouping patterns, showing that these loss patterns were not diagnostic of glaucoma. Further, we noted that 4 of 35 normal eyes exhibited a temporal wedge-shaped loss pattern. This suggests that loss in the temporal quadrant may be a less reliable diagnostic observation.

### 3.4 Characteristic of Nerve Fiber Layer Parameters

All NFL parameters, including the three reflectance and one thickness parameters were significantly different between the normal and glaucoma groups (Table 3). The overall average thickness and reflectance were normally distributed for all groups (Fig. 6). The low-reflectance superpixel count and focal reflectance loss parameters were not normally distributed. The normal group clustered around zero for both the low-reflectance superpixel count and the focal reflectance loss parameters. The PPG group had a trimodal distribution for the low-reflectance superpixel count, and a bimodal distribution for the focal reflectance loss. The PG group had a bimodal distribution for both the low-reflectance superpixel count and the focal reflectance loss. The different distribution patterns for average and focal parameters suggested that the glaucoma groups might not be homogeneous, and thus, there may be distinct clusters of focal versus diffuse loss patterns.

Unsupervised cluster analysis based on Gaussian mixture models[52] (Fig. 7) showed 3 loss patterns. In Cluster 1, most normal eyes (27/35) and 8 PPG eyes had no reflectance loss. In Cluster 2, eight normal eyes, 18 PPG, and 26 PG eyes had equal diffuse and focal losses. In Cluster 3, 4 PPG and 9 PG eyes had predominantly focal loss. Generally, Cluster 3 had a more severe average (p=0.044) and focal (p=0.001) reflectance loss than Cluster 2. This suggests that the predominantly focal pattern of loss may be associated more aggressive disease courses.

### 3.5 Diagnostic Accuracy

All NFL reflectance parameters had significantly higher AROCs than the average thickness (Table 4). Eyes with focal reflectance loss and low-reflectance pixel count had higher AROCs than did those with average reflectance, but the differences were not significant.

All reflectance parameters had significantly higher glaucoma diagnostic sensitivity than did thickness parameters when the specificity was fixed at 99% (Table 5). Focal reflectance loss had the highest overall sensitivity, detecting over half of the PPG eyes and nearly all of the PG eyes.

Using either the 5% or 1% cutoff, focal reflectance loss detected more glaucoma eyes than did the average NFL thickness (p≤0.01). Venn diagrams (Fig. 8) showed that nearly all eyes with abnormally thin NFL thicknesses also had abnormally large focal reflectance loss, but not vice versa. Thus NFL thickness would not be needed if focal reflectance loss was already used as the primary diagnostic parameter.

### 3.6 Correlation with Visual Field

All NFL parameters had moderate correlation with VF MD (Table 6). Focal reflectance loss had the highest correlation, but it was not significantly higher than the average NFL thickness. The NFL reflectance parameters were highly correlated with NFL thickness. All of the correlations with VF MD and NFL thickness were statistically highly significant (p < 0.001).

Two-segmented piecewise linear regression showed that all NFL reflectance and thickness parameters had good correlation with VF MD for eyes with no or mild VF loss (Fig. 9). However, they were poorly correlated for eyes with moderate to severe loss. This floor effect suggests that all NFL parameters are suitable for glaucoma monitoring in only the early stages.

## 4 Discussion

NFL reflectivity loss probably precedes thinning because the decrease of axonal microtubes occurs prior to loss of axons and NFL thinning.[28-30] Microtubule content can also be measured by birefringence measured by polarimetry or polarization-sensitive OCT.[53-56] Indeed, loss of NFL birefringence precedes thinning by 3 months in monkeys[57] and by 1 week in rats.[58] So theoretically these approaches could improve the early detection of glaucoma. However, clinical measurements of both NFL birefringence and reflectivity are very challenging because of many extrinsic factors that introduce noise and bias. For reflectivity measurements based on OCT, important extrinsic factors include beam coupling and incidence angle. The goal of our investigation and algorithm development effort was to reduce the effects of these extrinsic noises and more cleanly recover the diagnostic information in OCT scans of the peripapillary NFL.

Beam coupling refers to the efficiency with which the tissue reflection is coupled back to the OCT detection system. Coupling is reduced by defocus, astigmatism, higher-order aberrations, iris vignetting, media opacity (cataract, vitreous floaters), and polarization mismatch (corneal birefringence and other factors). Generally, variation in beam coupling is best compensated by the normalization of NFL reflectance against a reference layer that would be equally affected. We previously described normalization of NFL reflectivity by that of the retinal pigment epithelium (RPE), and found it improved glaucoma diagnostic accuracy.[31] Gardiner reported that normalization improved the repeatability of reflectivity measurements.[32] Liu et al. combined normalized NFL reflectivity with thickness to generate a reflectance index, and found it further improved diagnostic sensitivity in glaucoma suspects.[33] Our approach here was similar to that of Liu et al. because we integrated reflectivity over the NFL to produce a normalized reflectance. We made a slight change in that we expanded the reference layer to include the ellipsoid band as well as the RPE to improve robustness. A drawback to this approach is that peripapillary atrophy of the outer retinal layers could artifactually increase the normalized reflectance and interfere with the detection of NFL loss in these areas. However, previous studies and this study showed that overall this approach increased glaucoma diagnostic accuracy.

Incidence angle variation is a more subtle issue. Knighton et al. showed that reflectivity of nerve fibers is negatively related to the incident angle (with the angle defined as zero at perpendicular incidence), and the relationship is shaped like a Gaussian curve.[35] In OCT scanning, the incidence angle depends on the beam location in the pupil, the axial length, and the curvature of the retina. The OCT operator could adjust the positioning of the machine until the retinal cross section appears as flat as possible, thus reducing the variation of the incidence angle. However, this is difficult to achieve while avoiding iris vignetting and while keeping the retina within the image frame. The effect of incidence angle variation on NFL reflectance cannot be reduced by employing the RPE as a reference layer because RPE reflectivity is not similarly affected by incidence angle.[59, 60]

As far as we know, our method of azimuthal filtering is the first attempt to reduce the effect of incidence angle variation on NFL reflectance measurement. Our results showed that azimuthal filtering improved the repeatability of NFL reflectance measurement, reduced inter-individual variation among normal subjects, and improved glaucoma diagnostic accuracy. The main disadvantage of azimuthal filtering is the reduction of diagnostic information associated with asymmetric NFL loss in glaucoma. However, our results showed that overall the approach improved diagnostic accuracy. A more perfect solution would be to maintain perpendicular incidence while scanning the NFL, but none of the commercial OCT systems on the market has this functionality.

Another strategy that we successfully employed was the algorithm to measured focal NFL reflectance loss. Focal loss is measured in areas that have sufficiently severe loss that measurement noise is insignificant by comparison. Our results showed that this strategy further improved diagnostic accuracy. With the focal reflectance loss parameter, we were able to detect a majority of PPG eyes and almost all PG eyes at a specificity level of 99%. This is a major improvement over the NFL thickness parameter and may be sufficiently high to be useful in the population-based screening of at-risk patients. However, we cannot be sure that the excellent results we obtained here would fully generalize to populations with different characteristics. Even though we had used a cross-validation technique to reduce bias in our diagnostic accuracy assessment, our study population is different from the general population in that it had been selected to reduce confounding factors. In the general population, common pathologies such as epiretinal membrane, high refractive error, retinal edema, and retinal hemorrhage might interfere with reflectance analysis. Patients with other types of glaucoma may have different patterns of reflectance loss. Thus independent population-based studies would be needed to validate our findings.

An added bonus in our focal loss analysis is the emergence of a class of glaucoma patients in which focal loss predominates over diffuse loss. This cluster had significantly more severe disease in our study population, suggesting that disease progression in these patients may be more rapid. Thus focal NFL reflectance loss may be a valuable prognostic biomarker for the speed of glaucoma progression. This agrees with our previous results in the Advanced Imaging for Glaucoma study,[61] in which we found that focal loss in macular GCC and peripapillary NFL thickness were the best predictors of future VF progression.[14, 46, 62] We hypothesize that predominantly focal NFL reflectance loss may be an indication of a local defect in the structure or perfusion of the optic nerve head, similar to those found in eyes with disc hemorrhage, laminar defect, or peripapillary choroidal defect.[63-65] A longitudinal study is needed to assess this prognostic potential.

Beyond focal loss analysis, other patterns in the normalized NFL reflectance map may offer additional diagnostic information. We found that diffuse and wedge-shaped reflectance defects are characteristic of glaucoma, with the possible exception of temporal wedges. Our superpixel grid, which followed the trajectory of nerve fibers, facilitated the detection of the wedge patterns. These patterns could be automatically analyzed with machine learning methods, including deep learning. Indeed, other investigators have found deep learning to be useful in analyzing OCT images to detect glaucoma.[22, 66] The sample size of this study is too small to train a deep learning neural network, but the potential exists to apply this methodology to the analysis of normalized reflectance maps when a larger sample of clinical data becomes available. Our reflectance map is derived from a volumetric OCTA scan from which a capillary density map could also be obtained.[42] Thus it is possible to combine both reflectance and perfusion maps in the same pattern analysis.

A major limitation of NFL reflectance parameters is the presence of a floor effect. This limitation is well known for NFL thickness parameters. Both reflectance and thickness decrease with disease severity as measured by VF MD, but only in mild glaucoma. In moderate to severe glaucoma stages, both NFL reflectance and thickness reach a floor value that do not reflect further gradations. This means that NFL reflectance may be less useful in the staging and monitoring of glaucoma beyond the early onset of the disease. Fortunately, other objective measures of glaucoma, such as macular ganglion cell complex thickness[13] and OCT angiography perfusion measurements,[42, 67] may be better for this purpose.

## 5 Conclusions

We have shown that azimuthal filtering and focal loss analysis improves the glaucoma diagnostic value of NFL reflectance measurements to a level that is significantly higher than the widely used NFL thickness parameter. Subjects with predominantly focal rather than diffuse reflectance loss tend to have more severe glaucoma. Focal NFL reflectance loss is a promising OCT-derived diagnostic biomarker for the early detection of glaucoma and a prognostic biomarker to predict the rate of disease progression. However, due to the floor effect, NFL reflectance loss is only suitable for monitoring disease progression in the early stages.

**Author Contributions**

O.T. and D.H. designed the study. O.T. and D. H. wrote the manuscript and all coauthors critically commented and/or edited the manuscript. D.H. supervised the project. L.L. and Q.Y. did the manual grading. J.W. and Y.L. developed the Center for Ophthalmic Optics & Lasers-Angiography Reading Toolkit (COOL-ART) software. A.C, E, I and J.M. conduct the clinical study.

**Competing Interests Statement**

OHSU, Dr. Tan, Dr. Huang, and Dr. Jia have a significant financial interest in Optovue, Inc., a company that may have a commercial interest in the results of this research and technology. These potential conflicts of interest has been reviewed and managed by OHSU.

**Figures and Tables:**

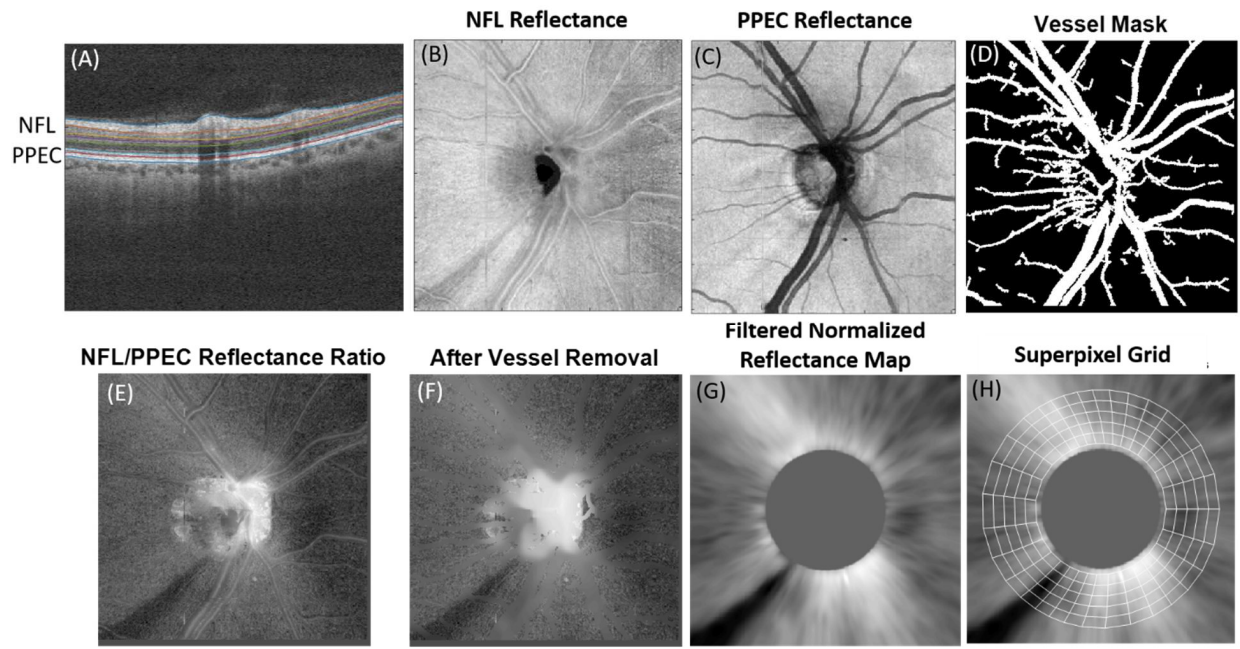

**Figure 1.** Calculation of the nerve fiber layer (NFL) reflectance map (4.5×4.5 mm) in a glaucomatous right eye with an inferotemporal nerve fiber bundle defect. (A) OCT sections were segmented to identify the topmost NFL and a reference layer called the photoreceptor-pigment epithelium complex (PPEC). (B) Summed OCT signal intensity map in the NFL band. (C) Average OCT signal intensity map in the PPEC band. (D) Large vessel mask. (E) NFL/PPEC reflectance ratio map. (F) Ratio map with vessels removed. (G) Formation of the normalized NFL reflectance map by normalization of the ratio map against the average value from the normal population and then performing spatial frequency filtering in the polar coordinate. (H) Reflectance map is overlaid with a superpixel grid.

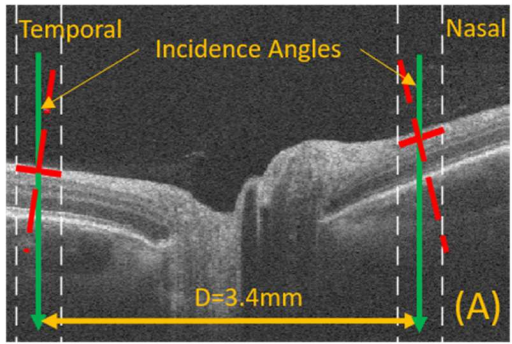
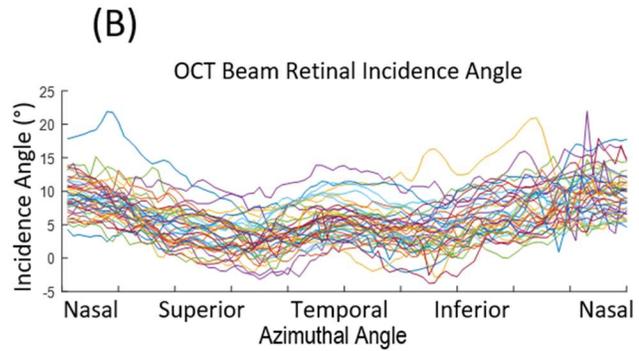
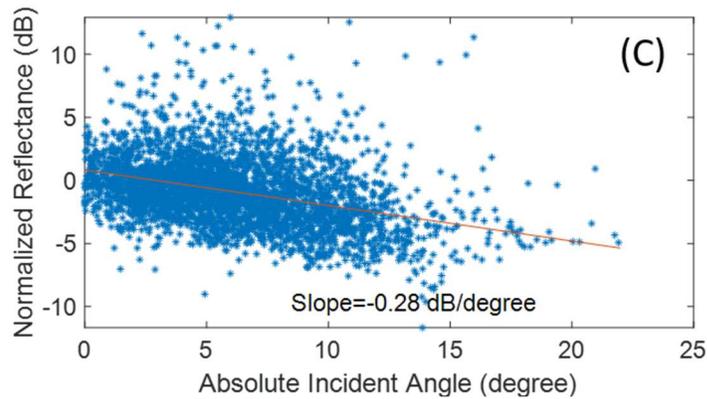

**Figure 2.** Azimuthal variation in incidence angle and NFL reflectance. (A) The radial OCT section across the optic nerve head shows that the beam incidence direction (green arrows) is usually slightly centripetal relative to the surface normal vector (red dashed lines) in the peripapillary retina. The incidence angle was measured by the radial slope (short red lines) of the inner retinal surface with perpendicular incidence defined as 0° and centripetal incidence defined as positive. (B) OCT beam incident angle along a 3.4-mm diameter circle in 35 normal eyes showed azimuthal sinusoidal variation with both first and second order periodicity. (C) Scatter plot of incident angle v. normalized reflectance (pooled superpixels on circle with dimeter 3.4 mm from 35 normal eyes). Linear regression showed that reflectance decreased with increasing incidence angle (p<0.001, r =-0.369).

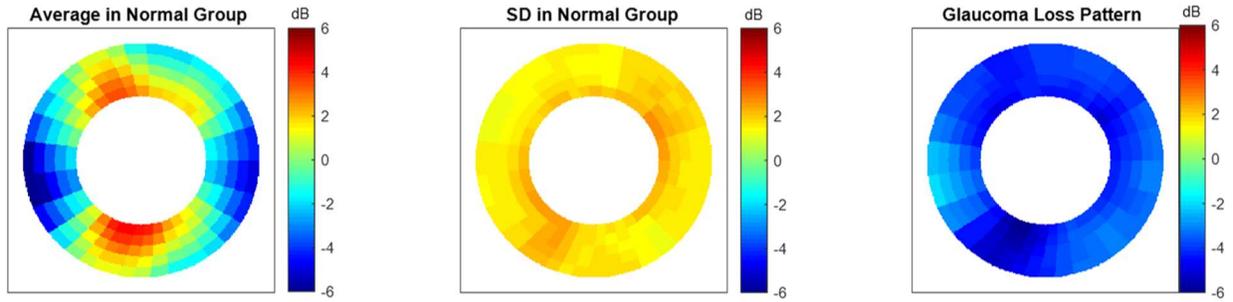

**Figure 3**. The normalized NFL reflectance maps averaged in the normal and glaucoma groups. The glaucoma group included both pre-perimetric and perimetric glaucoma cases. All eyes were transformed to a right-eye orientation for analysis. (Left) Average map of normal eyes. (Middle) The population standard deviation (SD) in the normal group. (Right) The average map for the glaucoma groups were subtracted by the normal average to obtain the average loss pattern (glaucoma damage shows as negative values).

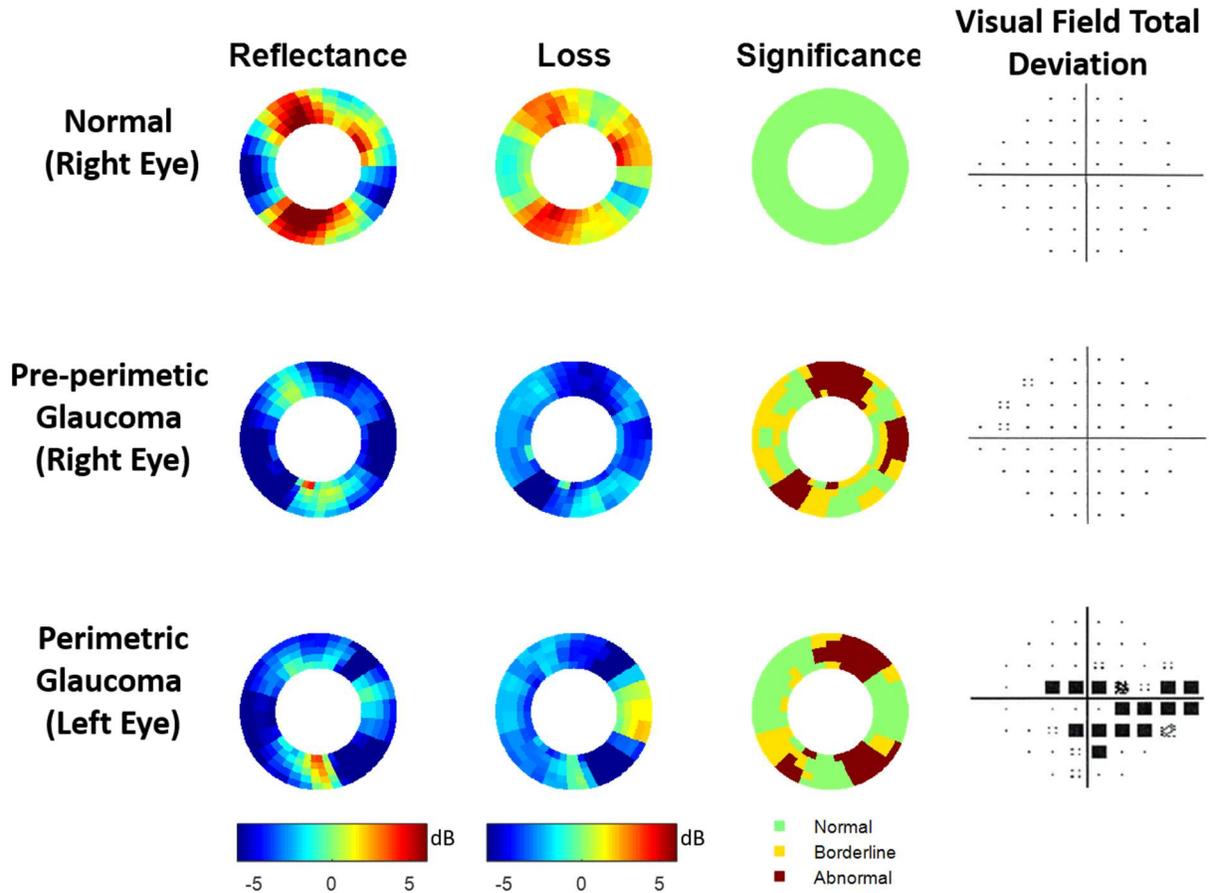

**Figure 4**. Representative NFL reflectance and VF maps from the normal and glaucoma groups. The significance map classifies superpixels into normal, borderline (1-5 percentile of normal population), and abnormal (below 1 percentile of normal) categories.

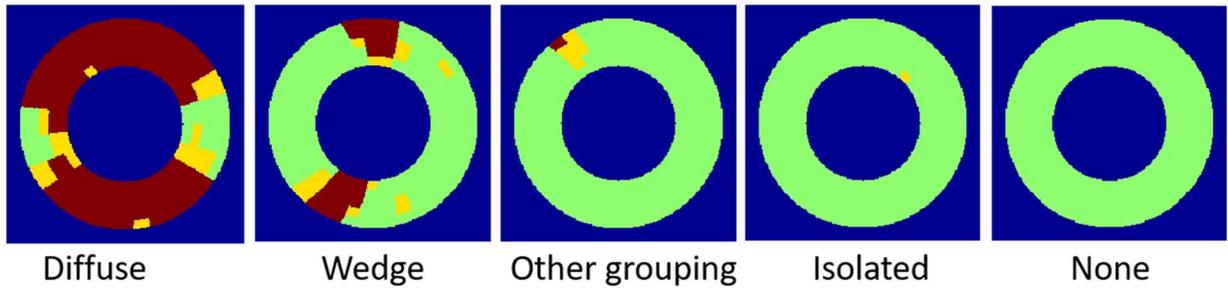
**Figure 5.** Five types of NFL reflectance loss patterns.

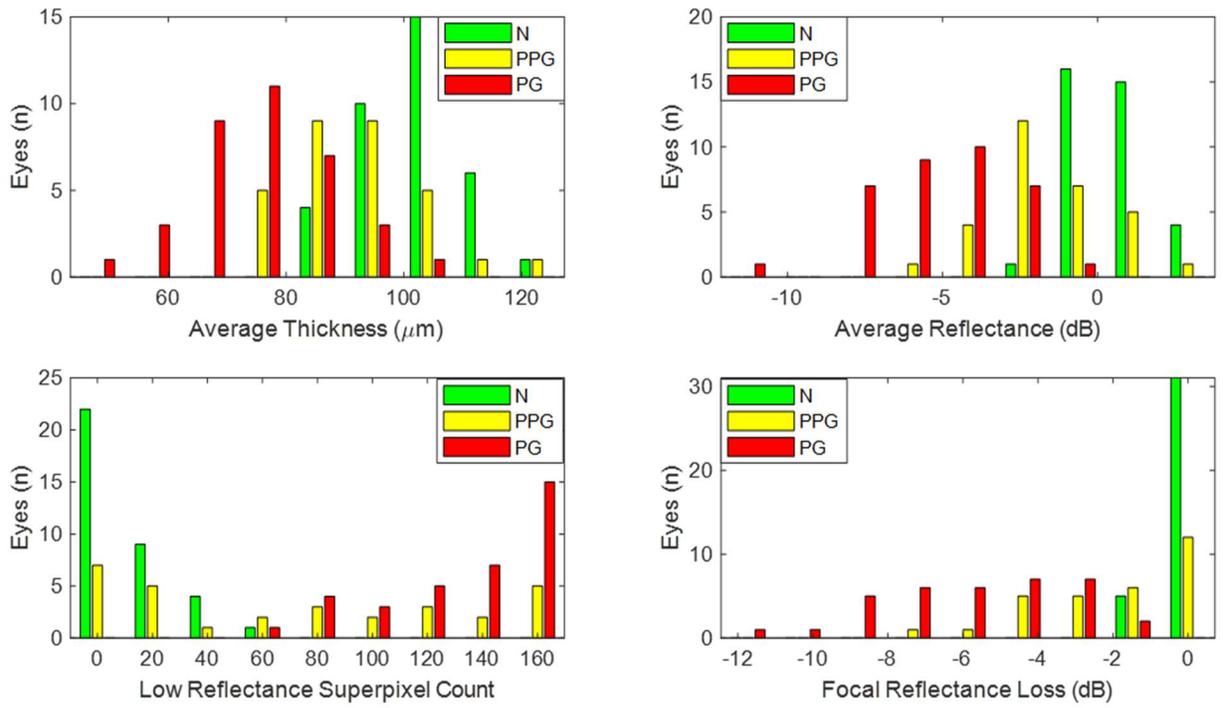

**Figure 6**. Distribution of nerve fiber layer parameters in three groups: normal (N), pre-perimetric glaucoma (PPG), and perimetric glaucoma (PG).

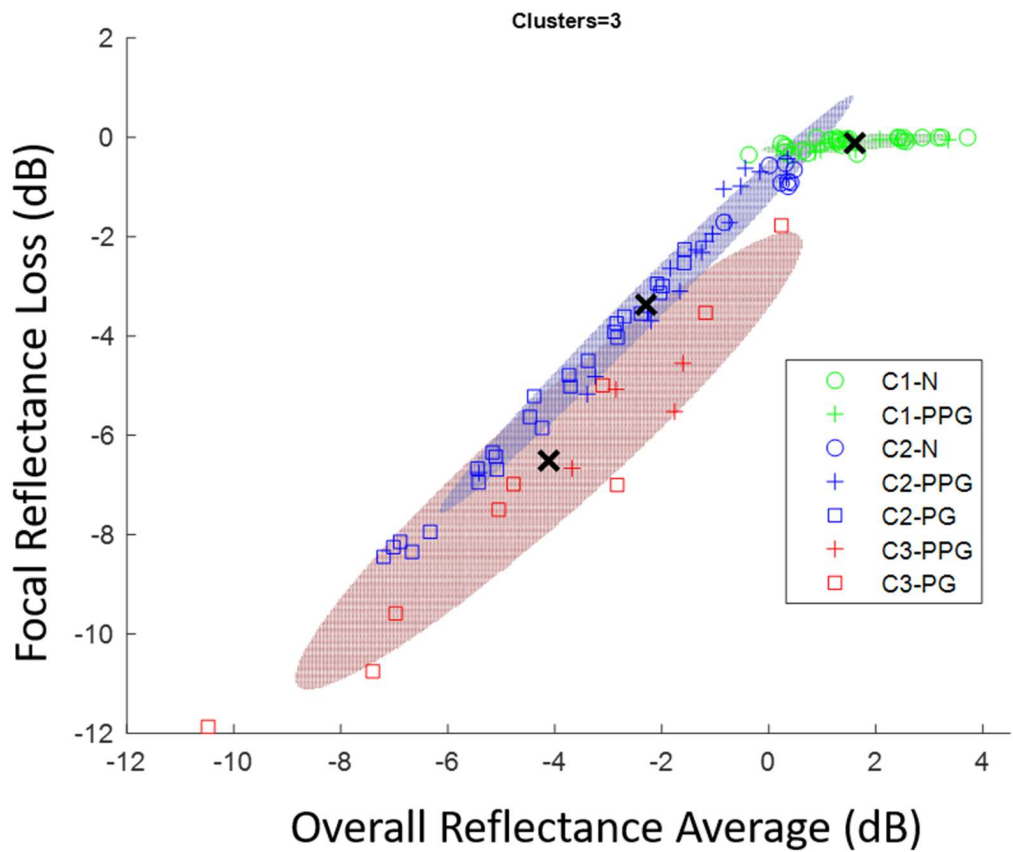

**Figure 7.** Unsupervised cluster analysis of focal versus overall reflectance loss revealed three clusters (C1 – C3): C1 = no loss (green); C2 = equal diffuse and focal loss (blue), and C3 = predominantly focal loss (red). These clusters were only partially correlated with the clinical diagnostic grouping: normal (circles), pre-perimetric glaucoma (cross), and perimetric glaucoma (square).

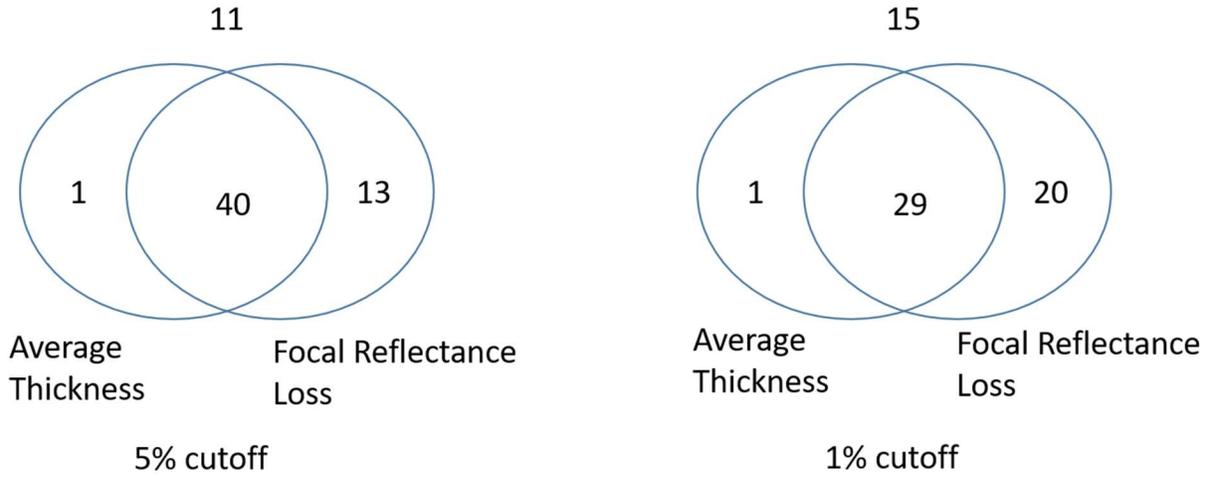

**Figure 8**. Venn diagrams of glaucoma detection with NFL parameters with either 5% or 1% specificity cutoff. Number in the circle are the eyes detected by either NFL parameter or both, while the number out of box is the eyes missed by both parameters. The PPG and PG groups were combined for this analysis. NFL = nerve fiber layer; PPG = pre-perimetric glaucoma; PG = perimetric glaucoma.

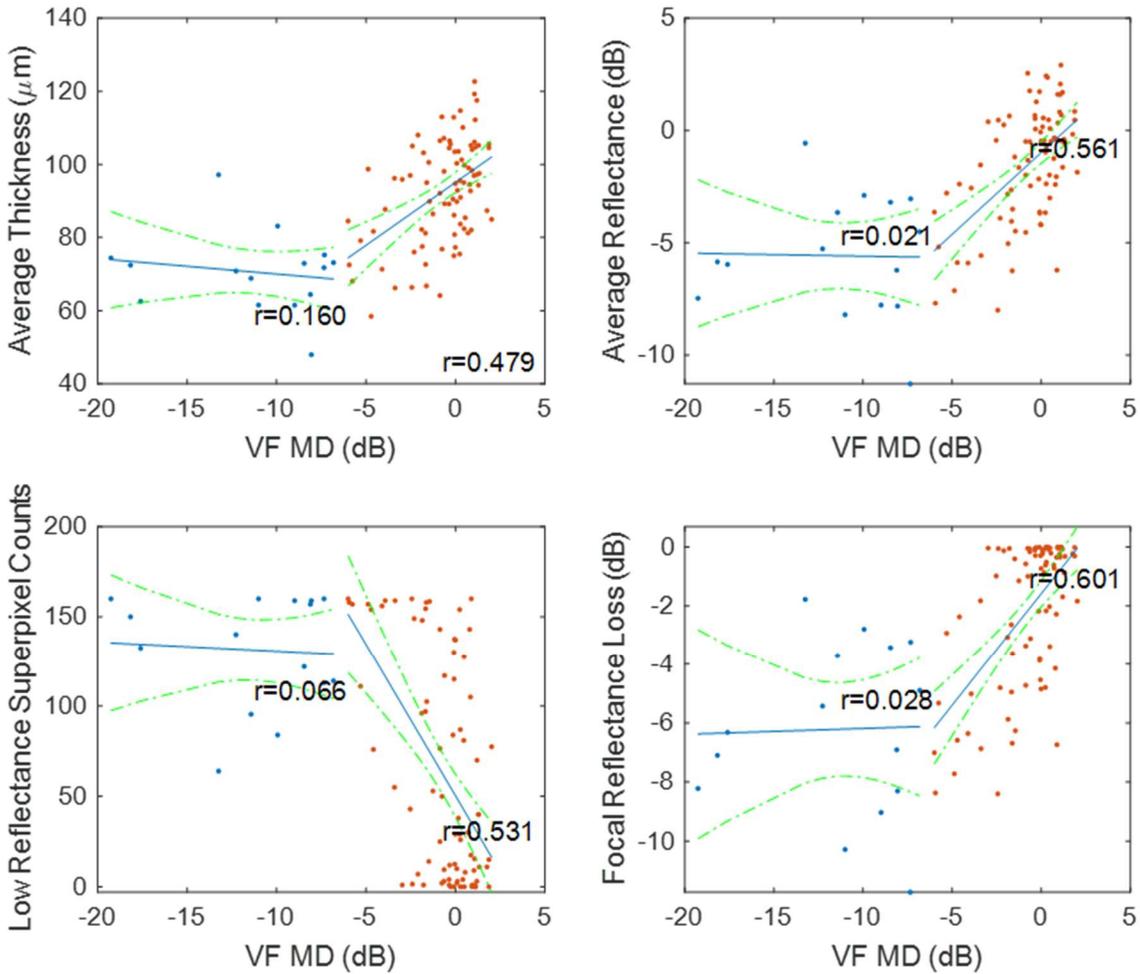

**Figure 9.** Two-segment piecewise linear regression of NFL parameters against VF MD. The plots combine normal, PPG, and PG groups. The segments for all NFL parameters in normal and early glaucoma eyes (VF MD > -6 dB), marked by red dots, were significantly correlated with VF MD (p < 0.001). The segments for moderate and severe glaucoma (VF MD < -6 dB), marked by blue dots, were not correlated with VF MD (p > 0.05). NFL = nerve fiber layer; VF = visual field; MD = mean deviation; PPG = pre-perimetric glaucoma; PG = perimetric glaucoma.

**Table 1. Characteristics of the Study Population**

|  | Normal | PPG | PG |
|---|---|---|---|
| Eye # | 35 | 30 | 35 |
| Age (Year) | 60.0 ± 10.8 | 65.1 ± 8.7* | 66.9 ± 8.8* |
| Gender (Male/Female) | 8/27 | 12/18* | 21/14* |
| Axial Length (mm) | 23.6 ± 0.9 | 24.7 ± 1.0* | 24.6 ± 1.3* |
| VF MD (dB) | 0.23 ± 1.24 | -0.63 ± 1.89* | -6.06 ± 5.20* |
| VF PSD (dB) | 1.46 ± 0.31 | 1.82 ± 0.63* | 7.29 ± 4.30* |

PPG = pre-perimetric glaucoma. PG = perimetric glaucoma. VF MD = visual field mean deviation. Values for continuous variables are means ± standard deviations. *, p-value < 0.05 compared to the normal group. VF PSD = visual field pattern standard deviation.

**Table 2. Eyes with Different Loss Pattern in Normal, PPG and PG Eyes**

| Defect pattern | Normal Eyes | PPG Eyes | PG Eyes |
|---|---|---|---|
| Diffuse | 0 | 10 | 24 |
| Wedge | 4 | 12 | 11 |
| Other grouping | 13 | 4 | 0 |
| Isolated | 6 | 3 | 0 |
| None | 12 | 1 | 0 |

PPG = pre-perimetric glaucoma; PG = perimetric glaucoma

**Table 3. Group Statistics for Nerve Fiber Layer Parameters**

|  |  | Normal | Glaucoma | p-value |
|---|---|---|---|---|
| Thickness | Overall average (µm) | 102.06±8.77 | 82.74±14.67 | <0.001 |
|  | Focal loss volume (%) | -0.85±2.35 | -14.07±12.81 | <0.001 |
|  | Inferior Quadrant(µm) | 127.74±14.37 | 96.45±23.18 | <0.001 |
| Reflectance | Average loss (dB) | 0.42±1.11 | -3.45±2.73 | <0.001 |
|  | Low-reflectance superpixel count | 11.51±16.11 | 106.27±54.70 | <0.001 |
|  | Focal Loss (dB) | -0.28±0.38 | -4.14±2.90 | <0.001 |

**Table 4. Diagnostic Accuracy of Nerve Fiber Layer Parameters**

| NFL | | AROC | Confidence Interval (95%) | p-value |
|---|---|---|---|---|
| Thickness | Overall average | 0.859±0.037 | 0.788,0.931 | N/A |
| | Focal loss volume | 0.861±0.032 | 0.799,0.923 | N/A |
| | Inferior quadrant | 0.862±0.036 | 0.792,0.931 | N/A |
| Reflectance | Overall average | 0.910±0.029 | 0.853,0.967 | 0.061 |
| | Low-reflectance superpixel count | 0.921±0.026 | 0.870,0.973 | 0.044 |
| | Focal loss | 0.925±0.025 | 0.876,0.974 | 0.043 |

NFL = nerve fiber layer; AROC = area under receiver operating characteristic curve; p-values = differences between NFL reflectance parameters and best single NFL thickness parameters.

**Table 5. Diagnostic Sensitivity of Nerve Fiber Layer Parameters at 99% Specificity**

| NFL | PPG | p-value | PG | p-value | All Glaucoma | p-value |
|---|---|---|---|---|---|---|
| Average Thickness | 0.233 | N/A | 0.714 | N/A | 0.492 | N/A |
| Thickness focal loss volume | 0.100 | N/A | 0.657 | N/A | 0.400 | N/A |
| Inferior Quadrant Thickness | 0.167 | N/A | 0.800 | N/A | 0.507 | N/A |
| Average Reflectance | 0.367 | 0.289 | 0.943 | 0.073 | 0.677 | 0.006 |
| Low Reflectance Superpixel Count | 0.500 | 0.043 | 0.971 | 0.041 | 0.754 | <0.001 |
| Focal Reflectance Loss | 0.533 | 0.027 | 1 | 0.023 | 0.769 | <0.001 |

NFL = nerve fiber layer; PPG = pre-perimetric glaucoma; PG = perimetric glaucoma; p-values = differences between NFL reflectance parameters and best single NFL thickness parameter.

Table 6. Pearson Correlation Matrix of OCT and Visual Field Diagnostic Parameters

| Pearson R | Average Reflectance (r) | Low-Reflectance Superpixel Count (r) | Focal Reflectance Loss (r) | Average NFL Thickness (r) |
| --- | --- | --- | --- | --- |
| **VF MD** | 0.593 | -0.519 | 0.612 | 0.560 |
| **NFL thickness** | 0.854 | -0.815 | 0.790 | N/A |

VF MD = visual field mean deviation; NFL = nerve fiber layer**; r=**Pearson correlation coefficient;